\documentstyle[preprint,aps]{revtex}

\begin{document}

\title {Quantization of generally covariant systems with extrinsic time}

\author{Rafael Ferraro{\footnote{Electronic address:
ferraro@iafe.uba.ar}}} 

\address{{\tighten
{\it Instituto de Astronom\'\i a y F\'\i sica del Espacio, Casilla de 
Correo 67 - Sucursal 28, 1428 Buenos Aires, Argentina\\ and Departamento 
de F\'\i sica, Facultad de Ciencias Exactas y Naturales, Universidad de 
Buenos Aires - Ciudad Universitaria, Pabell\' on I, 1428 Buenos Aires, 
Argentina\\}}}

\author{Daniel M. Sforza{\footnote{Electronic address:
sforza@iafe.uba.ar}}}

\address{{\tighten
{\it Instituto de Astronom\'\i a y F\'\i sica del Espacio,
Casilla de Correo 67 - Sucursal 28, 1428 Buenos Aires, Argentina\\}}} 

\maketitle

\begin{abstract}

{\tighten

A generally covariant system can be deparametrized by means of an
``extrinsic'' time, provided that the metric has a conformal ``temporal''
Killing vector and the potential exhibits a suitable behavior with respect
to it. The quantization of the system is performed by giving the well
ordered constraint operators which satisfy the algebra. The searching of
these operators is enlightened by the methods of the BRST formalism. 

}

\end{abstract}  
\vskip  1cm 


\narrowtext

\newpage

General relativity and quantum mechanics are the most important 
achievements of physics in this century. It seems essential to find a
quantum theory of gravity by embracing both theories in a consistent
one. However, despite the many efforts that have been made, that program
has not been sucessfully completed\cite{k1}.

One of the most difficult features is the problem of time\cite{k2}. In
quantum mechanics, time is an absolute parameter; it is not on an equal
footing with the other coordinates that turn out to be operators and 
observables. Instead, in general relativity ``time'' is merely an arbitrary 
label of a spatial hypersurface, and physically significant quantities are
independent of those labels: they are invariant under diffeomorfisms.
General relativity is an example of a parametrized system (a system whose
action is invariant under change of the integrating parameter). One can
obtain such a kind of system by starting from an action which does not
possess reparametrization invariance, and raising the time to the rank of a dynamical variable. So the original degrees of freedom and the time are left as 
functions of some physically irrelevant parameter. Time can be varied
independently of the other degrees of freedom when a constraint together
with the respective Lagrange multiplier are added. In this process, one
ends with a special feature: the Hamiltonian is constrained to vanish.   

Most efforts directed to quantize general relativity (or some
minisuperspace models) emphasize the analogy with the relativistic
particle\cite{prd,hk}. Actually, both systems have Hamiltonian constraints 
${\cal H}$ that are hyperbolic on the momenta. If the role of the squared mass is
played by a positive definite potential, then the analogy is complete in
the sense that time is hidden in configuration space. In fact, the positive definite potential guarantees that the temporal component of the momentum is never null on the constraint hypersurface. Thus the Poisson bracket $\{q^o, {\cal H} \}$ is also never
 null, telling us that $q^o$ evolves monotonically on any dynamical trajectory; this is the essential property of time. In this case, Ref.\cite{prd} shows the consistent operator ordering obtained from the Becchi-Rouet-Stora-Tyutin (BRST) formalism.

Unfortunately that analogy cannot be considered too seriously because the
potential in general relativity is the (non positive definite) spatial
curvature.  This means that the time in general relativity must be
suggested by another mechanical simile.

In order to essay a better mechanical model for general relativity, let us
start with a system of $n$ genuine degrees of freedom with a Hamiltonian
$h={1\over 2}g^{\mu\nu}p_{\mu}p_{\nu}+v(q^{\mu})$  and definite positive
metric $g^{\mu\nu}$. 

The dynamics of the system will not change if a function of $t$, namely
$-{t^2\over 2}$ is added to the Hamiltonian. So we write the action 
\begin{equation}
S=\int \left[p_{\mu}{dq^{\mu}\over dt} - h(q^{\mu},p_{\mu})+{t\over 2}^2\right] dt,~~~~~ {\mu}=1,...,n
\end{equation}

The system is  parametrized by regarding the integration variable $t$ as a
canonical variable whose conjugated momentum is (minus) the
hamiltonian. This last condition enters the action as a constraint ${\cal
H}=p_t+h-{t\over 2}^2$, and the action reads 
\begin{equation}
S[q^{\mu},p_{\mu},t,p_t,N]=\int\left[p_t{dt\over d{\tau}}+ p_{\mu}{dq^{\mu}\over
d{\tau}} - N \left( p_t+h(q^{\mu},p_{\mu})-{t\over 2}^2 \right)\right] d\tau
\label{act}\end{equation}
where $N$ is the Lagrange multiplier.

So far the constraint is parabolic in the momenta. However one can perform
the canonical transformation,
\begin{equation}
q^0=p_t ,~~~~~~~~~~~~~~~  p_0=-t \label{tc}
\end{equation}
that turns the constraint ${\cal H}$ into a hyperbolic function of the
momenta:
\begin{equation}
{\cal H}=q^0+h-{1\over 2}p_0^2=-{1\over 2}p_0^2+{1\over 2} g^{\mu\nu}
p_{\mu}p_{\nu}+v(q^{\mu})+q^0={1\over 2} {\cal G}^{rs} p_r p_s+{\cal
V}(q^r), \label{ham}\end{equation} 
with $r,s=0,1,...,n$. The metric components read ${\cal G}^{00}=-1$, ${\cal
G}^{0\nu}=0$, ${\cal G}^{\mu\nu}=g^{\mu\nu}$  and the potential is
${\cal V}(q^i)=v(q^{\mu})+q^0$. Then
${\cal G}^{rs}$ is a Lorentzian metric, as is the supermetric in the Arnowitt-Deser-Misner (ADM) formalism of general relativity. The constraint (\ref{ham}) describes a parametrized system with extrinsic (hidden in the phase space) time\cite{bf}, whose po
tential is not positive definite.

For a complete analogy with general relativity, the ``supermomenta
constraints'' can be introduced by adding $m$ degrees of freedom $q^a$.
Their spurious character is stated by $m$ linear and homogeneus constraints
$G_a\equiv\xi_a^r p_r$, where $\vec\xi_a$ are $m$ vectors fields tangent to the
coordinate curves associated with the $q^a$'s. These $m$ constraints $G_a$
can still be linearly combined
\begin{equation}
G_a \rightarrow G_{a'}= A_{a'}^{~a}(q)~G_a,~~~~~~~~~~~~{\rm det}~A\ne
0,\label{combl} \end{equation} 
to get an equivalent set of linear and homogeneous supermomenta
constraints. The set $({\cal H}, G_a)$ is first class.

Finally the dynamics of the system is obtained by varying the action
\begin{equation}
S[q^i,p_i,N,N^a]=\int \left[p_i{dq^i\over d{\tau}} - N {\cal H} - N^a G_a\right] d\tau,
~~~~~~~i=0,1,...,n+m 
\end{equation}
where $N^a$ are the Lagrangian multipliers corresponding to the constraints
$G_a$.

Such a system satisfies the following conditions [which can be read from
the Hamiltonian constraint, Eq. (\ref{ham})]:
\begin{eqnarray}
{\partial {\cal G}^{ij}\over \partial q^0}&& \approx 0\label{conda}\\
{\partial {\cal V}\over \partial q^0}&& = 1 \label{condb}
\end{eqnarray}
The symbol ``$\approx$'' means ``weakly equal'' (the equality is restricted
to the submanifold defined by the constraints $G_a\approx 0$) and it
replaces the ordinary equality because the metric has a nonphysical sector
which may depend on $q^0$.

The parametrization of the system, which is still visible in
Eq. (\ref{ham}) due to the special form of the potential and the components 
of the metric, can be masked by means of a general coordinate
transformation. However, the distinctive geometrical properties of the
system, namely Eqs. (\ref{conda}),(\ref{condb}), can be
written in a geometrical (i.e., coordinate independent) language, by using
 Lie derivatives. Thus, Eqs. (\ref{ham}) and (\ref{conda}),(\ref{condb})
tell us that there exists a weakly {\it unitary} temporal Killing vector
field satisfying
\begin{equation}
{\cal L}_{\vec\xi _0}{\bf {\cal G}}\approx 0, \label{condga}
\end{equation}
such that
\begin{equation}
{\cal L}_{\vec\xi _0}{\cal V} = 1. \label{condgb}
\end{equation}

Differing from those approaches where a hyperbolic constraint like
Eq. (\ref{ham}) is compared with the one of a relativistic particle, and the
parameter of the Killing vector is regarded as the time \cite{koxf}, in our
treatment the time ($t=-p_0$) is the dynamical variable {\it conjugated} to the
parameter of the Killing vector.

In order to quantize the theory, we must find well ordered first class
constraint operators satisfying the quantum constraint algebra,
\begin{equation}
[\hat{\cal H},\hat G_a]= \hat c_{0a}^0 \hat{\cal H}+\hat c_{0a}^b\hat G_b
\label{ccqn}\end{equation}
\begin{equation}
[\hat G_a,\hat G_b]=\hat C_{ab}^c(q) \hat G_c,
\label{clqn}\end{equation}
where the structure function $c_{0a}^b$ is linear in the momenta,
 $c_{0a}^b(q,p)=c_{0a}^{bj}(q) p_j$.

However, it is apparent that the potential ${\cal V}$ commute with the
linear constraints $G_a$ (it is gauge invariant) and as a consecuence the
structure function $\hat c_{0a}^0$ must vanish. The algebra
(\ref{ccqn}),(\ref{clqn}) with $\hat c_{0a}^0=0$ was already solved in
Ref.\cite{prd}. There, the Dirac constraint operators were obtained within
the framework of the BRST formalism:
\begin{equation}
\hat{\cal H}={1\over 2} f^{-{1\over 2}}\hat p_i {\cal G}^{ij}f \hat
p_jf^{-{1\over 2}}+ {i\over 2} f^{1\over 2} c_{oa}^{aj}\hat p_j f^{-{1\over 
2}} + {\cal V}
\label{vcuad} \end{equation}
and
\begin{equation}
\hat G_a= f^{1\over 2} \xi^i_a \hat p_i f^{-{1\over 2}},\label{vlin}
\end{equation}
where the function $f=f(q)$ satisfies 
\begin{equation}
C^b_{ab}=f^{-1}(f\xi^i_a)_{,i}={\rm div}_{\tilde\alpha}~\vec\xi_a,
\label{div}\end{equation}    
($\tilde\alpha$ is the volume ${\tilde\alpha}\equiv f~dq^0\wedge...\wedge
dq^{n+m}$).
\footnote{The $(n+m+1)$-form $\tilde\alpha$ solving Eq. (15) is a volume
in the configuration space ${\cal M}$: ${\tilde\alpha}\equiv
{\tilde E}^1\wedge...\wedge {\tilde E}^m \wedge {\tilde \omega}$, where
$\{\tilde E^a\}$ is the dual basis of $\{\vec\xi_a\}$ in $T^*_{||}{\cal M}$ 
(the ``longitudinal'' tangent space, and ${\tilde \omega}\ =\ \omega(y)\
dy^0\wedge...\wedge dy^{n}$ is a closed $n$ form where the $y^r$'s are
$n+1$ functions which are left invariant by the gauge transformations
generated by the linear constraints ($dy^r(\vec\xi_a)=0,~~\forall r,a$).
$\tilde\alpha$ is the volume induced by the constraints in the gauge orbit, 
times a (nonchosen) volume in the ``reduced" space. For a detailed
demonstration, see Ref.\cite{prd}.}

The Dirac constraint operators (\ref{vcuad}),(\ref{vlin}) were obtained
from the quantum BRST 
generator, the central object of the method. The BRST generator is a
fermionic real function in an extended phase space spanned by the original 
canonical pairs $(q^i, p_i)$ and by $m+1$ fermionic canonical pairs
$(\eta^a, {\cal P}_a)$ (one for each constraint).  The quantum BRST
generator is a nilpotent Hermitian operator which reads for the system
under consideration:
\begin{eqnarray}
\hat\Omega=&&{\hat\eta}^o \hat {\cal H} + {\hat\eta}^a \hat G_a + {1\over
2} {\hat\eta}^o{\hat\eta}^a {\hat c}_{oa}^b \hat {\cal P}_b + {1\over 2} 
{\hat\eta}^a{\hat\eta}^b {\hat C}_{ab}^c \hat {\cal P}_c \nonumber\\ 
=&&{\hat\eta}^o\left({1\over 2}  f^{-{1\over 2}} \hat p_i {\cal G}^{ij} f
\hat p_j  f^{-{1\over 2}} + {i\over 2} f^{1\over 2} c_{oa}^{aj}\hat p_j
f^{-{1\over 2}}+ {\cal V}\right)+ \hat \eta^a f^{1\over 2} \xi^i_a \hat p_i
f^{-{1\over 2}}\nonumber\\ &&+ {1\over 2} \hat \eta^o \hat \eta^a
(f^{1\over 2} c_{oa}^{bj}\hat p_j f^{-{1\over 2}} + f^{-{1\over 2}} \hat 
p_j c_{oa}^{bj} f^{1\over 2} ) \hat {\cal P}_b + {1\over 2} \hat \eta^a 
\hat \eta^b C_{ab}^c \hat {\cal P}_c 
\label{omeordv}\end{eqnarray} 
(in this $\eta-{\cal P}$ ordering, the Dirac constraint operators and the 
structure function operators can be directly read from it\cite{ht}).

In Ref.\cite{prd} we started with a pseudo-Riemannian metric and a constant 
potential (a relativistic particle in curved space). That system had the
property $c^0_{0a}=0$, which facilitated the search for the nilpotent BRST
generator. After that, a general (but positive definite) potential was
introduced by means of a unitary transformation of the BRST generator, and
$c^0_{0a}$ turned to be non null. This procedure gave to the constraint
operators the invariance under scaling of the superhamiltonian constraint 
(without having recourse to a curvature term).

In the present case the system is not a relativistic particle: the 
potential is not constant (and neither is it positive definite), and the
time is not hidden among the coordinates. However $c^0_{0a}$ is still null
due to the gauge invariance of the potential. Once again, the invariance
under scaling of the superhamiltonian will be introduced by performing a
unitary transformation in the extended space.

The scaling of the Hamiltonian constraint
\begin{equation}
{\cal H} \rightarrow  H = F ~{\cal H},~~~~~~~~~~~~~F>0 \label{escc}
\end{equation}
(then ${\cal G}^{ij} \rightarrow G^{ij}= F~ {\cal G}^{ij},~ {\cal V}
\rightarrow V= F~{\cal V}$) relaxes the geometrical properties of
$\vec\xi_0$:
\begin{eqnarray}
|\vec\xi_0|= 1~~ &&\rightarrow ~~|\vec\xi_0|= F^{-{1\over 2}}
\label{condg'0}\\  
{\cal L}_{\vec\xi _0}{\bf {\cal G}}\approx 0 ~~&&\rightarrow ~~ {\cal
L}_{\vec\xi_0} {\bf G} \approx C~ {\bf G}\label{condga'}\\
{\cal L}_{\vec\xi _0}{\cal V}=1~~ &&\rightarrow ~~{\cal L}_{\vec\xi _0}V= 
C~ V+|\vec\xi_0|^{-2}\label{condgb'} 
\end{eqnarray}
with $C(q)=\vec\xi_0 ({\rm ln} F)=-2\vec\xi_0 ({\rm ln} |\vec\xi_0|)$. Thus
$\vec\xi_0$ becomes a weakly nonunitary conformal Killing vector.

At the quantum level, the corresponding scaling operation can be
accomplished by performing the unitary transformation of the quantum BRST 
generator
\begin{equation}
\hat \Omega  \rightarrow  e^{i\hat  M}~\hat\Omega~ e^{-i\hat M},\label{tu} 
\end{equation}
with 
\begin{equation}
\hat M=[\hat{\cal P}_o~{\rm ln} |\vec\xi_0|~\hat\eta^o -\hat\eta^o ~{\rm ln} 
|\vec\xi_0|~\hat{\cal P}_o],~~~~~~~|\vec\xi_0|>0
\end{equation}
leading to a new Hermitian and nilpotent BRST generator,  
\begin{eqnarray}
\hat\Omega=&&{\hat\eta}^o\left({1\over 2} |\vec\xi_0|^{-1}
f^{-{1\over 2}}\hat p_i {\cal G}^{ij}f \hat p_j f^{-{1\over 2}}
|\vec\xi_0|^{-1} + {i\over 2} |\vec\xi_0|^{-1}  f^{1\over 2}
c_{oa}^{aj}\hat p_j f^{-{1\over 2}} |\vec\xi_0|^{-1}+
|\vec\xi_0|^{-2}{\cal V}\right)\nonumber\\
&&+ \hat\eta^a |\vec\xi_0| f^{1\over 2} \xi^i_a \hat p_i
f^{-{1\over 2}}|\vec\xi_0|^{-1}-2 \hat\eta^o\hat\eta^a \xi^i_a ({\rm ln}
|\vec\xi_0|)_{,i}\hat {\cal P}_o\nonumber\\ 
&& + {1\over 2} \hat \eta^o \hat \eta^a |\vec\xi_0|^{-1} (f^{1\over 2} 
c_{oa}^{bj} \hat p_j f^{-{1\over 2}} + f^{-{1\over 2}} \hat p_j c_{oa}^{bj} 
f^{1\over 2})|\vec\xi_0|^{-1} \hat {\cal P}_b + 
{1\over 2} \hat \eta^a \hat\eta^b C_{ab}^c \hat {\cal P}_c,
\label{omegaesc}\end{eqnarray} 
which corresponds to constraint operators satisfying the scaling invariance 
(so that $G^{ij}=|\vec\xi_0|^{-2}~{\cal G}^{ij}$, 
$V=|\vec\xi_0|^{-2}~{\cal V}$, and
$C_{oa}^{bj}=|\vec\xi_0|^{-2}~c_{oa}^{bj}$) 
\begin{equation}\hat H={1\over 2}|\vec\xi_0|^{-1}
f^{-{1\over 2}} \hat p_i |\vec\xi_0|^2 G^{ij}f \hat p_j f^{-{1\over
2}}|\vec\xi_0|^{-1}+ 
{i\over 2}|\vec\xi_0|  f^{1\over 2} C_{oa}^{aj}\hat p_j
f^{-{1\over 2}} |\vec\xi_0|^{-1} + V
,\label{vcuadq'v}\end{equation}
\begin{equation}\hat G_a=|\vec\xi_0| f^{1\over 2}
\xi^i_a\hat p_i f^{-{1\over 2}}|\vec\xi_0|^{-1},
\label{vlinq'}\end{equation} 
with the corresponding set of structure functions, 
\begin{equation}\hat C_{oa}^o=-2 \xi^i_a ({\rm ln}
|\vec\xi_0|)_{,i},\end{equation}
\begin{equation}\hat  C_{oa}^b = {1\over 2} \left( |\vec\xi_0| f^{1\over
2} C_{oa}^{bj}  \hat  p_j  f^{-{1\over  2}} |\vec\xi_0|^{-1} + 
|\vec\xi_0|^{-1} f^{-{1\over 2}} \hat p_j C_{oa}^{bj} f^{1\over 2}
|\vec\xi_0| \right),
\end{equation}
\begin{equation}\hat C_{ab}^c=C_{ab}^c,\label{ef}\end{equation} 
all of them properly ordered for satisfying the constraint algebra,
\begin{equation}[\hat  H,\hat  G_a]=\hat C_{oa}^o \hat H  +  \hat
C_{oa}^b(q,p) \hat G_b, \label{ccq'}\end{equation} 
\begin{equation}[\hat G_a,\hat G_b]=\hat C_{ab}^c(q) \hat G_c.
\label{clq'}\end{equation}

The quantization procedure is not completed without a physical inner
product where the spureous degree of freedom are frozen by means of gauge 
fixing conditions. 
In addition to the $m$ gauge conditions $\chi^a$ related to spatial
spureous degrees of freedom, one should deal with the reparametrization
invariance, which is associated with the inclusion of time among the
dynamical variables. This is an easy task, as long as one follows the
paramerization process exposed at the very beginning of the present
work. At the level of Eq. (\ref{act}), it is apparent that one should
insert a delta, $\delta(t-t_0)$ ($\{t-t_0,{\cal H}\}=1)$,  to regularize
the inner product, which means to take the inner product at a given time
$t_0$, 
\begin{equation}
(\varphi_1,\varphi_2)_{t_0}=\int dt dq\
\left[\prod^m \delta(\chi)\right]\ J\ \delta(t-t_0)\
\varphi^*_1(t,q^{\gamma})\ \varphi_2(t,q^{\gamma}),\end{equation}
where $\gamma=1, ..., n+m$ and $J$ is the Faddeev-Popov determinant
associated with the linear constraints.
Then, through a canonical transformation, the time is associated with the
momenta [Eq. (\ref{tc})]; so by changing to this representation (by
transforming Fourier the wave function) one obtains the inner product:
\begin{equation}
(\varphi_1,\varphi_2)={1\over 2\pi}\int dq\ dq^0\ dq'^0\ \left[\prod^m
\delta(\chi)\right]\ J\ e^{-it_0(q^0-q'^0)}\varphi^*_1(q^0,q^{\gamma})\
\varphi_2(q'^0,q^{\gamma}).\label{prodescreg}
\end{equation}

When the Hamiltonian constraint is scaled [Eq. (\ref{escc})], the physical
inner product must remain invariant. On account of the behavior of the wave 
function under scaling, which is apparent in the structure of the
constraint operators [see Eq. (\ref{wf}) below], the (originally unitary)
norm of $\vec\xi_0$ must appear in the physical inner product
\begin{equation}
(\varphi_1,\varphi_2)={1\over 2\pi}\int dq\ dq^0 \left[\prod^m
\delta(\chi)\right]\ J\ \varphi^*_1(q^0,q^{\gamma}) \ |\vec\xi_0|^{-1} \int
dq'^0\ e^{-it_0(q^0-q'^0)} |\vec\xi_0|^{-1}\varphi_2(q'^0,q^{\gamma}).
\label{pescrege}\end{equation}
It should be noticed that the integration in the coordinate $q'^0$ is
evaluated along the vector field lines of $\vec\xi_0$.

After completing the quantization, one can further understand the obtained
ordering, Eqs. (\ref{vcuadq'v}),(\ref{ef}). It fulfills the invariance
properties imposed to the theory: (i) coordinate changes, (ii) combinations 
of the supermomenta [Eq. (\ref{combl})], and (iii) scaling of the
super-Hamiltonian [Eq. (\ref{escc})]. The physical gauge-invariant inner
product of the Dirac wave functions, Eq. (\ref{pescrege}), must be
invariant under any of these transformations. On account of the change of
the Faddeev-Popov determinant under (ii) and (iii), the inner product will 
remain invariant if the Dirac wave function changes according to
\begin{equation}
\varphi  \rightarrow    \varphi'=({\rm det}    A)^{1\over    2}
|\vec\xi_0|~ \varphi. \label{wf}
\end{equation}
So, the factors $f^{{\pm}{1\over 2}}$, $|\vec\xi_0|^{{\pm}1}$ in the
constraint operators  are just what are
needed in order that $\hat G_a\varphi,~\hat H\varphi,$ and ${\hat
C}_{oa}^b\varphi$ transform as $\varphi$, so preserving the geometrical
character of the Dirac wave function\cite{prd}.

Concerning the extension of the here exposed treatment to general
relativity, Kucha\v{r} has shown in Ref. \cite{koxf} that a conformal
timelike Killing vector actually exits in the superspace of the ADM formalism.
But, the question whether or not it satisfies property (\ref{condgb'})
remains open.

As a final remark, it is worth mentioning that although the idea of an
extrinsic time is not new in general relativity\cite{york}, its use in the
quantization problems is rather scarce\cite{kt1,kt2}. As it was shown, the
mechanical model presented here can lead to a better understanding of its
implementation.

\acknowledgments

This research was supported by Universidad de Buenos Aires (Proy. TX 064)
and Consejo Nacional de Investigaciones Cient\'\i ficas y T\'ecnicas.

\end{document}